\documentclass[a4paper]{article}
\usepackage{amssymb}

\usepackage{graphicx}
\usepackage{amsmath}

\begin{document}

\title{Quantum state conversion between continuous variable and qubits systems }
\author{Xiao-yu Chen, Liang Han, Li-zhen Jiang \\
{\small \ Lab. of quantum information, China Institute of Metrology,
Hangzhou 310034, China}}
\date{}
\maketitle

\begin{abstract}
We investigate how quantum state can be converted between continuous
variable and qubits systems. Non-linear Jaynes-Cumings interaction
Hamiltonian is introduced to accomplish the conversion. Detail analysis on
the conversion of thermal state exhibits that pretty good fidelity can be
achieved.

\end{abstract}

\section{Introduction}

Quantum information processing (QIP) has been extensively studied for a
qubit system which is a quantum extension of a bit, spanning two-dimensional
Hilbert space. A qubit is realized by an electronic spin, a two-level atom,
the polarization of a photon and a superconductor among others. Parallelly,
much attentions have been paid to the QIP of quantum continuous variable
(CV) system, CV physical systems such as a harmonic oscillator, a rotator
and a light field are defined in infinitive-dimensional Hilbert space. In
contrast to the easily manipulation and storage of qubit system, CV system
is more suitable for quantum information transformation. Until recently,
qubit and CV systems are nearly always treated separately. The physical
interface between CV and qubit system is waiting to be exploited. There have
been some pilot works on how to convert quantum continuous variable to qubit
system, but the efficient of the conversion is quite low\cite{Paternostro}.
We would propose a scheme of converting the quantum information between the
two systems with high fidelity in this letter.

Quantum CV system is described by density operator $\rho $ which is a
function of the annihilation operators $a_1,\cdots ,a_s$ and creation
operators $a_1^{\dagger },\cdots ,a_s^{\dagger }$ satisfying canonical
communication relation $[a_j,a_k^{\dagger }]=\delta _{jk}.$The subscripts of
the annihilation or creation operator are for different modes (frequencies
and polarizations of the optical state). A qubit is described by density
operator of two-dimensional Hilbert space. Let the basis of the
two-dimensional Hilbert space be $\left| -\right\rangle $ and $\left|
+\right\rangle $ (i.e. lower and higher levers of two levels atomic system),
the density operator is $\varrho =\alpha \left| -\right\rangle \left\langle
-\right| +\beta \left| -\right\rangle \left\langle +\right| +\beta
^{*}\left| +\right\rangle \left\langle -\right| +(1-\alpha )\left|
+\right\rangle \left\langle +\right| $. In the following we will consider
the quantum information conversion between single mode CV field and qubits
system or vice via. The aim of the quantum state conversion is to convert CV
quantum state to qubits system with high fidelity (or the reverse process),
that is, we use many qubits to express the CV quantum state. It seems that
this is much like quantum source coding\cite{Schumacher}, but they are very
different. In quantum source coding of a CV state $\rho ^{\prime }$, the aim
is to convert $\rho ^{\prime \otimes m}$ (many copies of $\rho ^{\prime },$%
i.i.d) to $\varrho ^{\prime \otimes n}$ with least $n/m$ at high fidelity.
The qubit state $\varrho ^{\prime }$ in quantum source coding is a state
with entropy almost being equal to 1(thus the state $\rho ^{\prime }$ is
efficiently compressed). While in our quantum state conversion, the input
state $\rho $ is only one copy, and the serial of output states $\varrho _i$
$(i=1,\cdots ,n)$ may not be identical and usually not be maximal entropy
states.

\section{The scheme of quantum state conversion}

In quantum information theory, all logical operations on qubit as well as
transformation on CV should be implemented with unitary transformation. The
interaction among different systems gives rise to the transfer of quantum
information between the systems. The scheme considered here is one mode
field interact with each qubit of the qubit series consecutively, similar to
that used in quantum information dilution\cite{Ziman}. Let the length of the
qubit series be $K.$ The $k-th$ step of interaction has the form that the
field and the $k-th$ qubit are prepared in the product state $\rho
_k(0)\otimes \varrho _k(0),$ where $\varrho _k(0)$ is in the definite
initial state $\left| -\right\rangle \left\langle -\right| .$ The whole
system will evolve in the way of $\ U_k\left( t_k\right) \rho _k\left(
0\right) \otimes \varrho _k\left( 0\right) U_k^{+}\left( t_k\right) $ ,
where $U_k\left( t_k\right) =\exp (-\frac i\hbar H_kt_k)$ is the evolution
operator in interaction picture, with $H_k$ being the interaction
Hamiltonian which characterizes the interaction between the field and the $%
k-th$ qubit. After the evolution for a time interval $t_k,$ the field and
the qubit may get entangled. At the end of the time interval, the
interaction between the field and the $k-th$ qubit will be turned off. The
field density operator now is $\rho _k(t_k),$ we set it as the initial state
$\rho _{k+1}(0)$ of the next step, where time is reset to $0$. Then the
field is moved to interact with the $(k+1)-th$ qubit and leave the $k-th$
qubit alone. The former qubits should not be dropped. After the field
interact with each qubit we drop the field, that is, we trace off the field
density operator. The field state is transformed to that of a series of
qubit states. Although the qubits are prepared in the same definite initial
state $\left| -\right\rangle \left\langle -\right| ,$ and there are no
direct interaction among qubits, the final state of the qubit system may be
an entangled system via the field interacts with each qubit. The residue
field state will tend to the vacuum state after the conversion as $K$ tends
to infinitive. We will elucidate this by an example in the next section.

\section{Conversion of quantum thermal state}

For simplicity, we choose quantum thermal state to elucidate the whole idea
of quantum state conversion. The density operator of thermal state is $\rho =
$ $(1-v)\sum_{m=0}^\infty v^m\left| m\right\rangle \left\langle m\right| $,
where $v=N/(N+1),$ with $N$ being the average photon number of the state. $%
\left| m\right\rangle $ is the number state with photon number $m,$ it is
the eigenstate of particle number operator $\widehat{n}=a^{+}a$. Before the
first step of conversion, the combined state is $\rho \otimes \left|
-\right\rangle _{1,1}\left\langle -\right| ,$ the subscript $1$ represents
the first qubit. Suppose the interaction model Hamiltonian of CV field
system and the atomic qubit system is
\begin{equation}
H_1=\hbar \Omega \left( \sqrt{\widehat{n}}a^{+}\sigma _{-}+a\sqrt{\widehat{n}%
}\sigma _{+}\right) ,  \label{wave1}
\end{equation}
where $\sigma _{-}$ and $\sigma _{+}$ are operators which convert the atom
state form its excited state $\left| +\right\rangle $ to ground state $%
\left| -\right\rangle $ and from ground state to excited state respectively.
$\hbar \Omega $ is the energy difference of the two energy levels. The
Hamiltonian (\ref{wave1}) can be considered as a kind of nonlinear
Jaynes-Cummings model\cite{Gerry}\cite{Mista}. Then
\begin{equation}
\exp (-\frac i\hbar H_1t_1)\left| m,-\right\rangle _1=\cos (m\Omega
t_1)\left| m,-\right\rangle _1-i\sin (m\Omega t_1)\left| m-1,+\right\rangle
_1.
\end{equation}
If the interaction time $t_1$ is adjusted in such a way that $\Omega t_1=\pi
/2,$ then
\begin{eqnarray}
\exp (-\frac i\hbar H_1t_1)\left| 2m,-\right\rangle _1 &=&(-1)^m\left|
2m,-\right\rangle _1, \\
\exp (-\frac i\hbar H_1t_1)\left| 2m+1,-\right\rangle _1 &=&-i(-1)^m\left|
2m,+\right\rangle _1.  \nonumber
\end{eqnarray}
Applying the evolution operator $U_1(t_1)=\exp (-\frac i\hbar H_1t_1)$ to
the state $\rho \otimes \left| -\right\rangle _{1,1}\left\langle -\right| $,
we have
\begin{equation}
U_1(t_1)\rho \otimes \left| -\right\rangle _{1,1}\left\langle -\right|
U_1^{\dagger }(t_1)=(1-v)\sum_{m=0}^\infty v^{2m}\left| 2m\right\rangle
\left\langle 2m\right| \otimes (\left| -\right\rangle _{1,1}\left\langle
-\right| +v\left| +\right\rangle _{1,1}\left\langle +\right| )
\end{equation}
It should be noticed that the state after evolution is a product state of CV
system state and qubit state, the field and the qubit become separable. Some
of the quantum information has been already transferred to qubit system,
this can be seen from the entropy of the qubit state. The first qubit state
after the transfer is
\begin{equation}
\varrho _1=\frac 1{1+v}(\left| -\right\rangle _{1,1}\left\langle -\right|
+v\left| +\right\rangle _{1,1}\left\langle +\right| ),
\end{equation}
whose entropy is $S(\varrho _1)=\log _2(1+v)-\frac v{1+v}\log _2v$. The
density operator of the field after the evolution will be
\begin{equation}
\rho _2(0)=\rho _1(t_1)=(1-v^2)\sum_{m=0}^\infty v^{2m}\left|
2m\right\rangle \left\langle 2m\right| .
\end{equation}
The CV state $\rho _2(0)$ has even number of photons in each item. We can
separate the first qubit state $\varrho _1$ from the combined state, then
append the second qubit state which is prepared in definite initial state $%
\left| -\right\rangle _{2,2}\left\langle -\right| $ to the field, the new
combined state will be $\rho _2(0)\otimes $ $\left| -\right\rangle
_{2,2}\left\langle -\right| $. We would design another interaction
Hamiltonian to assign part of CV state to the second qubit. The Hamiltonian
will be
\begin{equation}
H_2=\hbar \Omega (\sqrt{\widehat{n}}a^{+}\frac 1{\sqrt{\widehat{n}}%
}a^{+}\sigma _{-}+a\frac 1{\sqrt{\widehat{n}}}a\sqrt{\widehat{n}}\sigma
_{+}),
\end{equation}
the evolution operator will be $U_2(t_2)=\exp (-\frac i\hbar H_2t_2).$ Then
\begin{equation}
\exp [-\frac i\hbar H_2t_2]\left| 2m,-\right\rangle _2=\cos (2m\Omega
t_2)\left| 2m,-\right\rangle _2-i\sin (2m\Omega t_2)\left|
2m-2,+\right\rangle _2.
\end{equation}
If the interaction time $t_2$ is adjusted in such a way that $\Omega t_2=\pi
/4,$ then
\begin{eqnarray}
\exp (-\frac i\hbar H_2t_2)\left| 4m,-\right\rangle _2 &=&(-1)^m\left|
4m,-\right\rangle _2, \\
\exp (-\frac i\hbar H_2t_2)\left| 4m+2,-\right\rangle _2 &=&-i(-1)^m\left|
4m,+\right\rangle _2.  \nonumber
\end{eqnarray}
Applying the evolution operator $U_2(t_2)=\exp (-\frac i\hbar H_2t_2)$ to
the state $\rho _2(0)\otimes \left| -\right\rangle _{2,2}\left\langle
-\right| $, we have
\begin{equation}
U_2(t_2)\rho _2(0)\otimes \left| -\right\rangle _{2,2}\left\langle -\right|
U_2^{\dagger }(t_2)=(1-v^2)\sum_{m=0}^\infty v^{4m}\left| 4m\right\rangle
\left\langle 4m\right| \otimes (\left| -\right\rangle _{2,2}\left\langle
-\right| +v^2\left| +\right\rangle _{2,2}\left\langle +\right| )
\end{equation}
The state after evolution is a product state of CV system state and qubit
state again. The second qubit state after the transfer is
\begin{equation}
\varrho _2=\frac 1{1+v^2}(\left| -\right\rangle _{2,2}\left\langle -\right|
+v^2\left| +\right\rangle _{2,2}\left\langle +\right| ),
\end{equation}
whose entropy is $S(\varrho _2)=\log _2(1+v^2)-\frac{v^2}{1+v^2}\log _2v^2$.
The CV\ system after the evolution will be
\begin{equation}
\rho _3(0)=\rho _2(t_2)=(1-v^4)\sum_{m=0}^\infty v^{4m}\left|
4m\right\rangle \left\langle 4m\right| .
\end{equation}
The second qubit state $\varrho _2$ then is removed from the combined state,
and third qubit state is appended onto the field. The $k-th$ Hamiltonian
will be $H_k=\hbar \Omega [\widehat{n}(\frac 1{\sqrt{\widehat{n}}%
}a^{+})^{2^{k-1}}\sigma _{-}+(a\frac 1{\sqrt{\widehat{n}}})^{2^{k-1}}%
\widehat{n}\sigma _{+}]$ and interaction time is $t_k=\pi /(2^k\Omega ).$
After all the evolution the whole state will be
\begin{eqnarray}
&&U_K(t_K)\cdots U_2(t_2)U_1(t_1)\rho \otimes \left| -,-,\cdots
,-\right\rangle \left\langle -,-,\cdots ,-\right| U_1^{\dagger
}(t_1)U_2^{\dagger }(t_2)\cdots U_K^{\dagger }(t_K) \\
&=&(1-v^{2^K})\sum_{m=0}^\infty v^{2^Km}\left| 2^Km\right\rangle
\left\langle 2^Km\right| \otimes \varrho ,  \nonumber
\end{eqnarray}
where $\varrho =\varrho _1\otimes \varrho _2\otimes \cdots \otimes \varrho _K
$ with $\varrho _k=(\left| -\right\rangle _{k,k}\left\langle -\right|
+v^{2^{k-1}}\left| +\right\rangle _{k,k}\left\langle +\right|
)/(1+v^{2^{k-1}}).$ The information transferred to $k-th$ qubit is $%
S(\varrho _k)=\log _2(1+v^{2^{k-1}})-\frac{v^{2^{k-1}}}{1+v^{2^{k-1}}}\log
_2v^{2^{k-1}}$. The total information transferred is
\begin{equation}
\sum_{k=1}^KS\left( \varrho _k\right) =S(\rho )-S[\rho _K(t_K)].
\end{equation}
The residue state of the field after all the evolution is $\rho
_K(t_K)=(1-v^{2^K})\sum_{m=0}^\infty v^{2^Km}\left| 2^Km\right\rangle
\left\langle 2^Km\right| ,$ with its entropy being
\begin{equation}
S[\rho _K(t_K)]=-\log _2(1-v^{2^K})-\frac{2^Kv^{2^K}}{1-v^{2^K}}\log _2v.
\end{equation}
The residue CV\ state then is traced. The entropy loss is $S[\rho _K(t_K)]$
in this quantum state conversion procedure. In each step of evolution the
entropy of the combined state is preserved, this is because unitary
operation does not change the entropy of the combined state. The entropy
loss can only occur at the last step of dropping the residue CV\ system.
When $K\rightarrow \infty ,$ we have $S[\rho _K(t_K)]\rightarrow 0,$ the
entropy transferred to the qubit system tends to $S(\rho )$. The information
is perfectly transferred. While for finite $K,$ from the qubit series state $%
\varrho _1,\varrho _2,\cdots ,\varrho _K$, the reconstructed CV state (see
Section 5) is
\begin{equation}
\rho _r=\frac{1-v}{1-v^{2^K}}\sum_{m=0}^{2^K-1}v^m\left| m\right\rangle
\left\langle m\right| .
\end{equation}
The fidelity of the state conversion is
\begin{equation}
F=Tr(\sqrt{\sqrt{\rho }\rho _r\sqrt{\rho }})=\sqrt{1-\nu ^{2^K}}.
\end{equation}
Other measure of the state conversion fidelity is the closeness of the
residue field to the vacuum filed state, which is
\begin{equation}
F^{\prime }=\sqrt{\left\langle 0\right| \rho _K(t_K)\left| 0\right\rangle }=%
\sqrt{1-\nu ^{2^K}}.
\end{equation}
The two fidelities are equal in quantum thermal state conversion. For most
of the thermal states ($\nu \nrightarrow 1$), pretty good fidelity can be
achieved by using several qubits.

\section{General quantum state conversion}

In this section we will consider quantum state conversion of a general
single mode quantum state. The initial CV state is $\rho
=\sum_{n,m=0}^\infty c_{nm}\left| n\right\rangle \left\langle m\right| ,$
the first step of conversion will be
\begin{eqnarray}
\exp [-\frac i\hbar H_1t_1]\rho \otimes \left| -\right\rangle
_{1,1}\left\langle -\right| \exp [\frac i\hbar H_1t_1]
&=&\sum_{n,m=0}^\infty (-1)^{m+n}\left| 2n\right\rangle \left\langle
2m\right| (c_{2n,2m}\left| -\right\rangle _{1,1}\left\langle -\right|
+ic_{2n,2m+1}\left| -\right\rangle _{1,1}\left\langle +\right|  \nonumber \\
&&-ic_{2n+1,2m}\left| +\right\rangle _{1,1}\left\langle -\right|
+c_{2n+1,2m+1}\left| +\right\rangle _{1,1}\left\langle +\right| ).
\end{eqnarray}
Then $\left| -\right\rangle _{2,2}\left\langle -\right| $ is appended and
the unitary transformation $\exp [-\frac i\hbar H_2t_2]$ is applied and so
on, at last $\left| -\right\rangle _{K,K}\left\langle -\right| $ is appended
and $\exp [-\frac i\hbar H_Kt_K]$ is applied, each item of the CV part will
convert to a form of $\left| 2^Kn\right\rangle \left\langle 2^Km\right| $.
At this stage the entropy of the combined state remains intact. We obtain
the qubit series by tracing the residue field. Denote $%
j=2^Kn+2^{K-1}j_K+2^{K-2}j_{K-1}+\cdots +j_1=2^Kn+(j_Kj_{K-1}\cdots j_1),$
with $j_k=0,1;$and $\left| -\right\rangle _k=\left| j_k=0\right\rangle
,\left| +\right\rangle _k=\left| j_k=1\right\rangle $ , then
\begin{eqnarray}
&&U_K(t_K)\cdots U_2(t_2)U_1(t_1)\rho \otimes \left| -,-,\cdots
,-\right\rangle \left\langle -,-,\cdots ,-\right| U_1^{\dagger
}(t_1)U_2^{\dagger }(t_2)\cdots U_K^{\dagger }(t_K) \\
&=&\sum_{n,m=0}^\infty (-1)^{m+n}\left| 2^Kn\right\rangle \left\langle
2^Km\right| \otimes \sum_{j_K,j_{K-1},\cdots ,j_1,l_K,l_{K-1},\cdots
,l_1=0,}^1c_{2^Kn+(j_Kj_{K-1}\cdots j_1)},_{2^Km+(l_Kl_{K-1}\cdots l_1)}
\nonumber \\
&&(-1)^{j_1+l_1}i^{j_K+j_{K-1}+\cdots +j_1}(-i)^{l_K+l_{K-1}+\cdots
+l_1}\left| j_Kj_{K-1}\cdots j_1\right\rangle \left\langle l_Kl_{K-1}\cdots
l_1\right| .  \nonumber
\end{eqnarray}
After tracing the residue field, the result $K-$qubit state will be
\begin{eqnarray}
\varrho &=&\sum_{m=0}^\infty \sum_{j_K,j_{K-1},\cdots
,j_1,l_K,l_{K-1},\cdots ,l_1=0,}^1c_{2^Km+(j_Kj_{K-1}\cdots
j_1)},_{2^Km+(l_Kl_{K-1}\cdots l_1)} \\
&&(-1)^{j_1+l_1}i^{j_K+j_{K-1}+\cdots +j_1}(-i)^{l_K+l_{K-1}+\cdots
+l_1}\left| j_Kj_{K-1}\cdots j_1\right\rangle \left\langle l_Kl_{K-1}\cdots
l_1\right| .  \nonumber
\end{eqnarray}
By the reverse conversion (see the next Section), the reconstruct field
state will be $\rho _r=\sum_{n,m=0}^{2^K-1}c_{nm}^{\prime }\left|
n\right\rangle \left\langle m\right| ,$ with $c_{nm}^{\prime
}=\sum_{n^{\prime },m^{\prime }=0}^\infty c_{2^Kn^{\prime }+n,2^Km^{\prime
}+m}$. When $K\rightarrow \infty ,$ we have $c_{nm}^{\prime }\rightarrow
c_{nm}$, the fidelity $F\rightarrow 1$. The fidelity may not be easily
calculated for a general input field state. Alternatively, we can calculate
the closeness of the residue field ($\rho _{residue}$) with respect to the
vacuum state, which turns out to be $F^{\prime }=\sqrt{\left\langle 0\right|
\rho _{residue}\left| 0\right\rangle }=\sqrt{\sum_{j=0}^{2^K-1}c_{jj}}.$ The
residue field state is obtained by tracing all qubit freedoms of the last
stage composed state. When the initial CV state is a coherent state $\left|
\alpha \right\rangle =\exp (-\left| \alpha \right| ^2/2)\alpha ^n/\sqrt{n!}%
\left| n\right\rangle $ , we have $F^{\prime }=\sqrt{\exp (-\left| \alpha
\right| ^2)\sum_{j=0}^{2^K-1}\left| \alpha \right| ^{2j}/j!}.$

\section{Reverse conversion}

In the reverse conversion, we have the initial state $\varrho _1\otimes
\varrho _2\otimes \cdots \otimes \varrho _K$, where $\varrho _k$ is the
density operator of $k-th$ qubit. The process of reverse state conversion is
to convert firstly the highest qubit $(K-th)$ to the CV state then the
lower. The combined state will evolve to
\begin{equation}
U_1^{\dagger }(t_1)\{\{U_2^{\dagger }(t_2)\cdots \{U_K^{\dagger }(t_K)\left|
0\right\rangle \left\langle 0\right| \otimes \varrho _KU_K(t_K)\}\cdots
\otimes \varrho _2U_2(t_2)\otimes \varrho _1\}U_1(t_1),
\end{equation}
where $\left| 0\right\rangle \left\langle 0\right| $ is the initial state of
the field. The first step is to transfer the state $\varrho _K=$ $\alpha
_K\left| -\right\rangle _{K,K}\left\langle -\right| +\beta _K\left|
-\right\rangle _{K,K}\left\langle +\right| +\beta _K^{*}\left|
+\right\rangle _{K,K}\left\langle -\right| +(1-\alpha _K)\left|
+\right\rangle _{K,K}\left\langle +\right| $ to the field. Since
\begin{eqnarray}
\exp (\frac i\hbar H_Kt_K)\left| 0,-\right\rangle _K &=&\left|
0,-\right\rangle _K \\
\exp (\frac i\hbar H_Kt_K)\left| 0,+\right\rangle _K &=&\cos (2^{K-1}\Omega
t_K)\left| 0,+\right\rangle _K+i\sin (2^{K-1}\Omega t_K)\left|
2^{K-1},-\right\rangle _K.  \nonumber
\end{eqnarray}
The evolution time is so chosen that $\cos (2^{K-1}\Omega t_K)=0,$ we choose
$2^{K-1}\Omega t_K=\pi /2$ as before. Then $\exp (\frac i\hbar H_Kt_K)\left|
0,+\right\rangle _K=$ $i\left| 2^{K-1},-\right\rangle _K$. The first step
evolution will be$U_K^{\dagger }(t_K)\varrho _K\otimes \left| 0\right\rangle
\left\langle 0\right| U_K(t_K)=\rho _K\otimes \left| -\right\rangle
_{K,K}\left\langle -\right| ,$ with
\begin{equation}
\rho _K=\alpha _K\left| 0\right\rangle \left\langle 0\right| -i\beta
_K\left| 0\right\rangle \left\langle 2^{K-1}\right| +i\beta _K^{*}\left|
2^{K-1}\right\rangle \left\langle 0\right| +(1-\alpha _K)\left|
2^{K-1}\right\rangle \left\langle 2^{K-1}\right|
\end{equation}
We see that all the information of qubit state $\varrho _K$ is transferred
to the field, leave the qubit state a definite blank state. Moreover, the
combined state is a direct product of the field and qubit state, thus the $%
K-th$ qubit can be dropped after the evolution. The next step is to transfer
$\varrho _{K-1}\ $to the remained field $\rho _K$. Since when $2^{K-2}\Omega
t_{K-1}=\pi /2,$ we have $U_{K-1}^{\dagger }(t_{K-1})\left| 0,-\right\rangle
_{K-1}=\left| 0,-\right\rangle _{K-1},$ $U_{K-1}^{\dagger }(t_{K-1})\left|
2^{K-1},-\right\rangle _{K-1}=-\left| 2^{K-1},-\right\rangle _{K-1},$ $%
U_{K-1}^{\dagger }(t_{K-1})\left| 0,+\right\rangle _{K-1}=$ $i\left|
2^{K-2},-\right\rangle _{K-1},$ $U_{K-1}^{\dagger }(t_{K-1})\left|
2^{K-1},+\right\rangle _{K-1}$ $=$ $-i\left| 2^{K-1}+2^{K-2},-\right\rangle
_{K-1}$. The state after the evolution will be $\rho _{K-1}\otimes \left|
-\right\rangle _{K-1,K-1}\left\langle -\right| \otimes \left| -\right\rangle
_{K,K}\left\langle -\right| .$ The quantum information of the two qubits are
transferred to $\rho _{K-1}.$ When all the qubits are transferred to the
field, we get at last a quantum CV state $\rho =\rho _1$ while leaving all
the qubit series in the lower energy level state $\left| -,-,\cdots
,-\right\rangle .$ Thus the reverse conversion procedure will convert a
general qubit pair product state $\varrho _1\otimes \varrho _2\otimes \cdots
\otimes \varrho _K$ into a continuous variable state $\rho $ while keeping
the entropy of the whole state. The reverse conversion is perfect.

\section{Conclusions}

The scheme of quantum state conversion between continuous variable and qubit
systems has been proposed based on the unitary evolution. The interaction
Hamiltonian is not a usual atom-photon (or electron-photon) interaction
Hamiltonian. It is a kind of nonlinear Jaynes-Cummings model. Although it is
physical, the realization of such a conversion is rather complicate and
difficult. Theoretically, the conversion is asymptotically perfect. After
the conversion, the field state is very close to vacuum, thus the field
state is transferred to the state of qubit series. The scheme of state
conversion can also be extended to the entanglement conversion between qubit
system and continuous variable system\cite{Chen}.

\section*{Acknowledgement}

Funding by the National Natural Science Foundation of China (under Grant No.
10575092,10347119), Zhejiang Province Natural Science Foundation (under
Grant No. RC104265) and AQSIQ of China (under Grant No. 2004QK38) are
gratefully acknowledged.

\end{document}